\documentclass[prc,showpacs,twocolumn]{revtex4}
\usepackage{graphicx}
\usepackage{dcolumn}
\usepackage{bm}

\begin{document}

\title{Microscopic description of fission in  
neutron-rich plutonium isotopes 
with the Gogny-D1M energy density functional}

\author{R. Rodr\'{\i}guez-Guzm\'an}

\email{raynerrobertorodriguez@gmail.com}

\affiliation{Department of  Physics and Astronomy, Rice University, 
Houston, Texas 77005, USA}

\affiliation{Department of Chemistry, Rice University, Houston, Texas 77005, USA}

\author{L.M. Robledo}

\affiliation{Departamento  de F\'{\i}sica Te\'orica, 
Universidad Aut\'onoma de Madrid, 28049-Madrid, Spain}

\email{luis.robledo@uam.es}

\date{\today}

\begin{abstract}
The most recent parametrization D1M of the Gogny energy density functional 
is used to describe fission in the isotopes $^{232-280}$Pu. We resort to the 
methodology introduced in our previous studies [Phys. Rev. C \textbf{88}, 054325 (2013)
and  Phys. Rev. C \textbf{89}, 054310 (2014)] to compute  the fission
paths, collective masses and zero point quantum corrections within the 
Hartree-Fock-Bogoliubov framework. The systematics of the spontaneous fission 
half-lives t$_{SF}$, masses and charges of the  fragments in Plutonium isotopes
is analyzed and compared with available experimental data. We also pay attention to 
isomeric states, the deformation properties of the  
fragments as well as to the competition between the spontaneous fission and 
$\alpha$-decay modes. The impact of pairing 
correlations on  
the predicted t$_{SF}$ values
 is demonstrated with the help of calculations 
for  $^{232-280}$Pu in which the pairing strengths of the Gogny-D1M energy
density functional are modified by 5 $\%$ and 10 $\%$, respectively. 
We further validate the use of 
the D1M parametrization  through the  discussion of  
the  half-lives in $^{242-262}$Fm. Our calculations 
corroborate
 that, though
the uncertainties in the absolute values of physical observables 
are large, the Gogny-D1M 
Hartree-Fock-Bogoliubov framework still reproduces the trends with mass and/or neutron
numbers and therefore represents a reasonable starting point to describe fission
in heavy nuclear systems from a microscopic point of view.

\end{abstract}


\maketitle

\section{Introduction}
Nuclear fission is a large amplitude collective phenomenon whose full
understanding still remains as one of the main challenges
in nuclear structure physics. On the way to scission into two or more fragments, the 
nuclear shapes evolve through a multidimensional landscape  
that can be described in terms of several deformation 
parameters \cite{Specht,Bjor,Boca-Raton-ref,Krappe}. Within this context, our 
present knowledge of nuclear fission owes a lot to the efforts 
to incorporate the stabilizing role of shells effects into the semiclasical
liquid drop model  (see, for example, Refs. \cite{Moller-1,Moller-2}
and references therein). The potential energy surfaces provided by such models
 emphasize the key role played by several kinds of nuclear 
configurations, intimately related to shells effects, along the fission path 
to determine observables like the 
spontaneous  fission half-life and the mass distribution of the 
fragments \cite{Moller-1}.
Such configurations comprise minima, valleys, ridges and saddle points. For example, local 
minima  can affect the 
dynamics and time scale of the fission process. In particular, both 
(superdeformed) first and (hyperdeformed) second  isomeric  states
have
been the subject of intense debate
\cite{Refs-barriers-other-nuclei-1,Pask,Moller-Nucl-Phys-1972,Kowal-thir-min,Berger-thir-min,Rutz-thir-min,Cwiok-third-min,Ben-third-min,Delaroche-2006,Mcdonell-2,Robledo-Giulliani,Rayner-Robledo-fission-U}.

%
%
\begin{figure}
\includegraphics[width=0.48\textwidth]{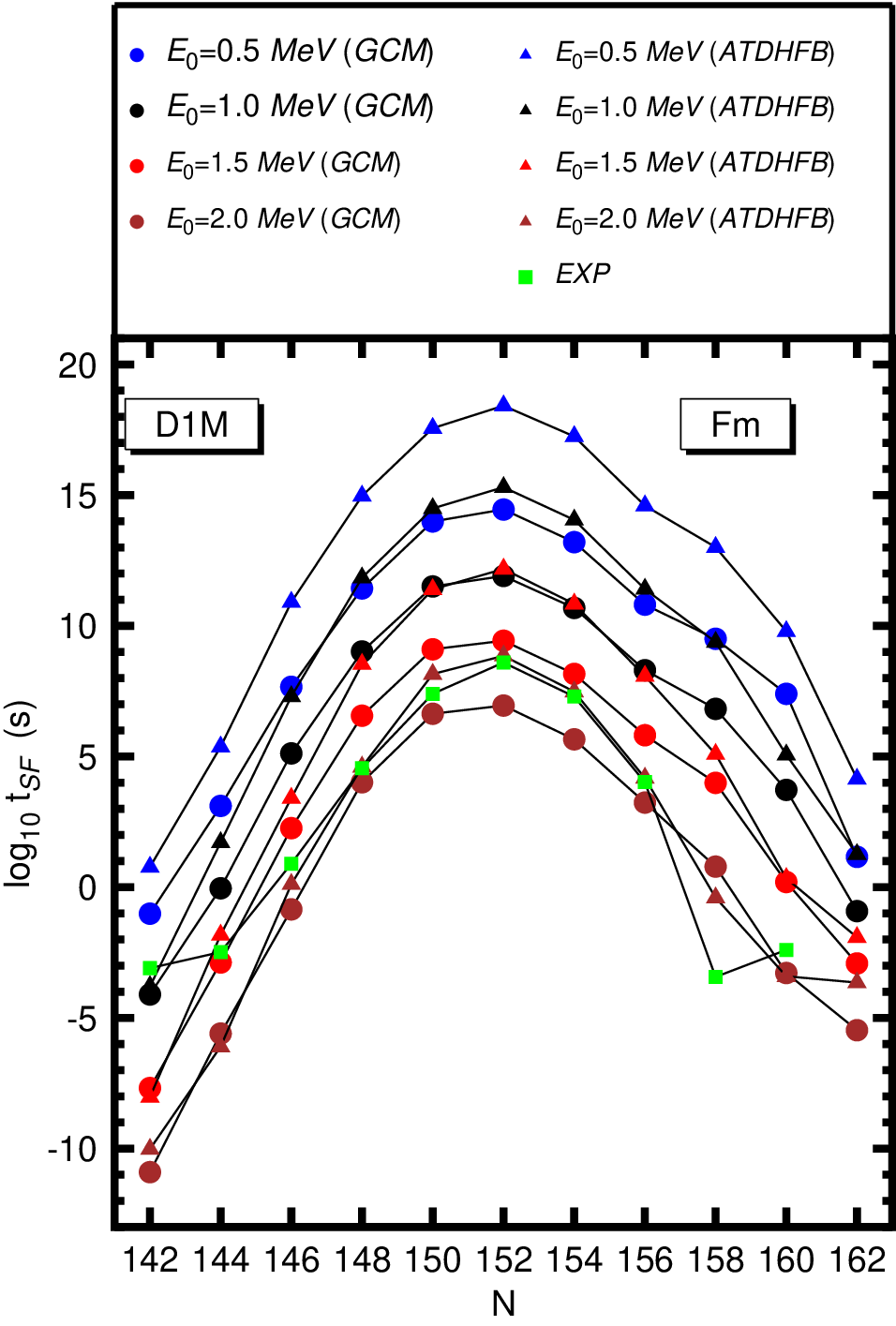}
\caption{ (Color online) The spontaneous fission half-lives t$_{SF}$, 
predicted within the GCM and ATDHFB schemes, 
for the isotopes $^{242-262}$Fm  are 
 depicted as functions of the neutron number. Results have been obtained with 
 the  Gogny-D1M EDF. Calculations have been carried out  with 
 E$_{0}$=0.5, 1.0, 1.5 and 
 2.0 MeV, respectively. The available experimental $t_{SF}$ values 
 \cite{Refs-barriers-other-nuclei-3-tsf}
 are included in the plot. For more details, see the main text.
}
\label{fig-Fm} 
\end{figure}

Though quite sophisticated approximations have already been invoked 
\cite{Negele,instanstons}, the constrained
 self-consistent mean-field approximation 
\cite{rs,Bender-review} has emerged as a poweful 
framework  for systematic microscopic studies of real fissioning nuclei
in terms of non-relativistic Gogny \cite{Delaroche-2006,Rayner-Robledo-fission-U,gogny-d1s,Robledo-Martin,Dubray,PEREZ-ROBLEDO,Younes2009,Warda-Egido-Robledo-Pomorski-2002,Warda-Egido-2012}, Skyrme \cite{Mcdonell-2,UNEDF1,Erler2012,Baran-1981,WN-Nature} and BCPM-like 
\cite{Robledo-Giulliani}
as well as 
relativistic \cite{Abusara-2010,Abu-2012-bheights,RMF-LU-2012,Kara-RMF} 
Energy Density Functionals (EDFs). Here, the multidimensional fission landscape is determined in terms of constraints 
on multipole moments and neck parameters. The approximation also provides the 
required ingredients to obtain the collective inertias as well as the zero
point energy quantum corrections \cite{Rayner-Robledo-fission-U} stemming from the restoration of the 
symmetries broken in the corresponding mean-field states through the 
spontaneous symmetry breaking mechanism \cite{rs}. 
It also accounts for quantum mechanical tunneling effects.
Such microscopic 
studies assume that fission properties are determined 
by general features of the considered EDFs and are 
quite demaning from the computational point of view, a task
that has been greatly  helped by recent 
developments in the field of high-performance computing.

From the theoretical point of view, a better description of the 
fission process is required to  account for shell 
effects and/or  magic numbers in heavy and superheavy nuclear 
systems \cite{WN-Nature,Sobiczewski}. On the other hand, microscopic studies of the
spontaneous fission and $\alpha$-decay modes \cite{Warda-Egido-2012,Erler2012}
are  important to better understand the stability of heavy and superheavy
elements. The last ones have been the subject of intense experimental
effort in recent years (see, for example, Refs. \cite{JULIN-SHE,Oganessian-3,Haba-SHE}
and references therein). Beside the unique insight that such 
superheavy elements provide on nuclear structure properties under
extreme conditions \cite{WN-Nature}, one shoud also keep in mind that they 
 are produced during the r-process and their properties
determine the upper end of the nucleosynthesis flow
\cite{Arnould-2007}. The wealth  of information in actinide nuclei \cite{Specht} 
as well as progress in several areas of science and applications \cite{Boca-Raton-ref,Krappe} also
act as driving forces to improve our models for nuclear fission. In addition, microscopic 
fission studies of neutron-rich nuclei are also required since, on the one hand, these are the 
territories where the fate of the nucleosynthesis of heavy nuclei is determined and, on the other
hand, such systems represent a challenging testing ground to examine the  
adequacy of nuclear effective interactions when extrapolated to exotic N/Z ratios.

%
%
\begin{figure}
\includegraphics[width=0.48\textwidth]{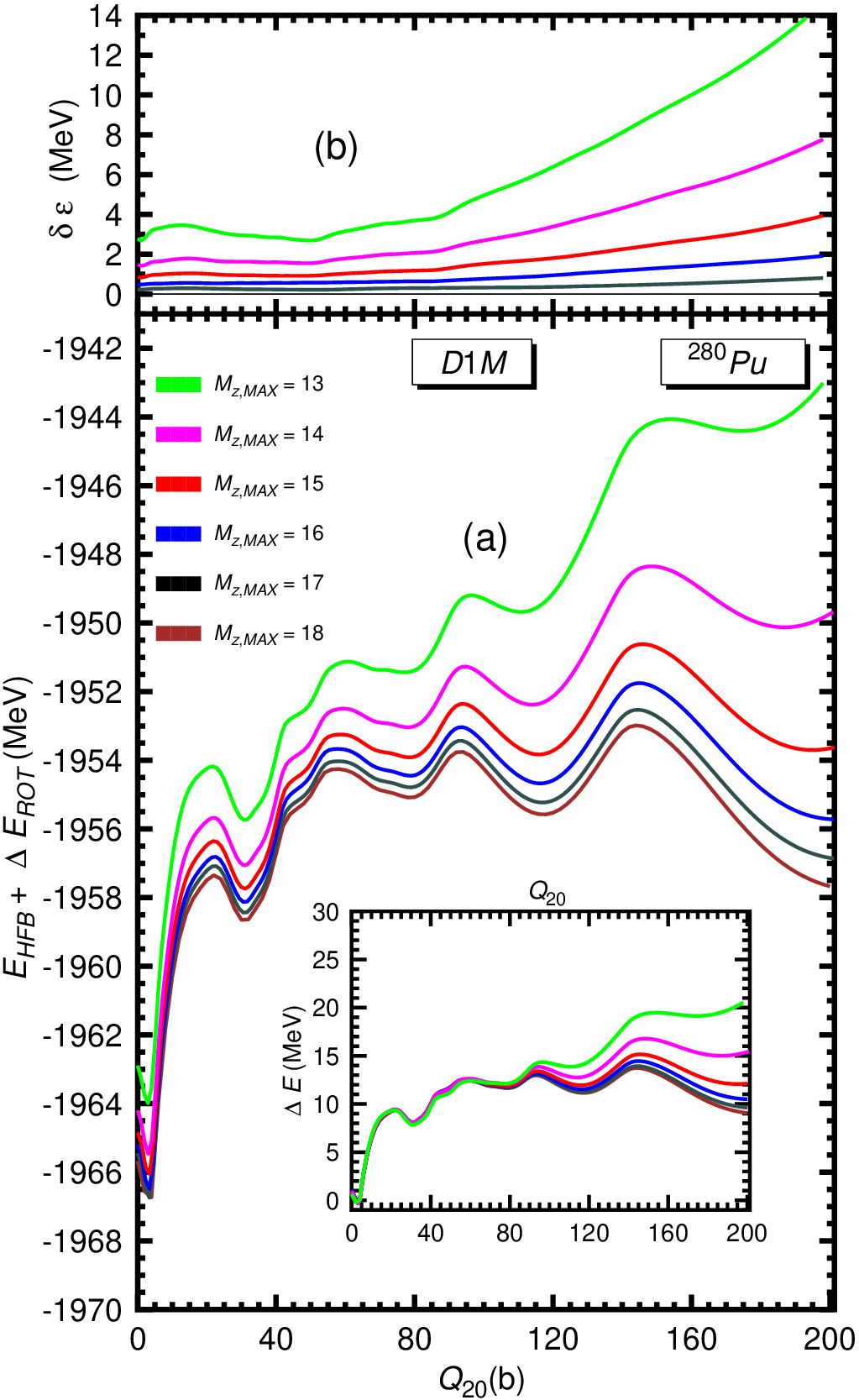}
\caption{ (Color online) The rotationally corrected energies
E$_{HFB}$ + $\Delta$ E$_{ROT}$ corresponding to the 1F configurations in 
$^{280}$Pu are shown  in panel (a) as functions of the quadrupole moment $Q_{20}$.
Results are shown for $M_{z,MAX}$=13, 14, 15, 16, 17 and 18, respectively.
The inset in panel (a) displays, for each $M_{z,MAX}$, the relative energies 
referred to the corresponding ground states. The energy differences with
respect to the calculations with  $M_{z,MAX}$=18 are depicted in panel (b).
For more details, see the main text.
}
\label{fig-convergence} 
\end{figure}

In our previous work \cite{Rayner-Robledo-fission-U}, we have 
performed dripline-to-dripline fission calculations for Uranium isotopes
as well as for a selected set of heavy and superheavy nuclei for which 
experimental data are available
\cite{Refs-barriers-other-nuclei-1,Refs-barriers-other-nuclei-2,Refs-barriers-other-nuclei-3-tsf,Pu-mass-fragments-exp-1,Pu-mass-fragments-exp-2}
. We have carried out a detailed comparison between the results obtained with the most standard parametrization of the
 Gogny-EDF \cite{Gogny-1980} (i.e., D1S \cite{gogny-d1s}) and the ones provided by the new
 parametrizations D1N \cite{gogny-d1n} and D1M \cite{gogny-d1m}, respectively. The comparison
 between Gogny-like EDFs,  with available 
 data for barrier heights, excitation energies of fission isomers and half-lives as well as with previous  
 theoretical studies \cite{Delaroche-2006,Robledo-Giulliani,Warda-Egido-Robledo-Pomorski-2002,Warda-Egido-2012}
have shown that the Gogny-D1M EDF represents a reasonable starting point to describe fission in 
heavy and superheavy nuclei. This is quite satisfying as the parametrization D1M 
does a much better job to reproduce nuclear masses \cite{gogny-d1m} and, at the same time, 
seems to reproduce low energy nuclear structure data with the same or better accuracy 
than the well tested D1S parametrization 
\cite{gogny-d1m,PRCQ2Q3-2012,Robledo-Rayner-JPG-2012,PTpaper-Rayner,Rayner-Sara,Rayner-Robledo-JPG-2009,Rayner-PRC-2010,Rayner-PRC-2011}. 
We have also paid special attention to the uncertainties in the determination of the absolute values 
of fission observables \cite{Robledo-Giulliani,Rayner-Robledo-fission-U}. Such uncertainties are presumed to be large. However, it has 
been shown that the mean-field approximation reproduces reasonably well the trend of fission observables 
as functions of the mass number and/or along isotoptic chains.

In the present work we have used the Hartree-Fock-Bogoliubov (HFB) approximation \cite{rs}, based on the Gogny-D1M EDF \cite{gogny-d1m}, to 
carry out fission calculations along the Plutonium isotopic chain, including very neutron-rich
isotopes. To this end, we have considered the 
nuclei $^{232-280}$Pu. We have used the same methodology as in Ref. \cite{Rayner-Robledo-fission-U}. 
Therefore, all the calculations to be discussed later on are subject to the same 
uncertainties already described in that
reference. However this study is, to the best of our knowlegde, the first one in which the Gogny-D1M EDF is 
systematically
employed  to describe fission in neutron-rich Plutonium isotopes. Second, it 
will allow us to see to which extent the physical trends already obtained for Uranium isotopes \cite{Rayner-Robledo-fission-U}, using this 
particular version of the Gogny-EDF, hold for other 
nuclei in the same region of the nuclear chart. Third, we discuss the appearence of second 
fission isomers in the corresponding 
one-fragment 
 curves of Plutonium nuclei.
Having in mind that fission observables 
are quite sensitive to pairing correlations \cite{Robledo-Giulliani,Rayner-Robledo-fission-U}, due 
to the strong dependence of the collective inertias with the inverse of the 
pairing gap \cite{proportional-1,proportional-2} , we have also carried out self-consistent HFB calculations 
for the nuclei $^{232-280}$Pu
using a 
modified Gogny-D1M EDF in which the strenghts of the neutron and proton pairing fields
are increased by 5 $\%$ and 10 $\%$, respectively. We also pay attention to 
the competition between the spontaneous fission and $\alpha$-decay channels
along the Plutonium  chain. Last, but not least, we further validate the use of the Gogny-D1M EDF
in fission studies, by comparing the spontaneous fission half-lives
predicted for the nuclei $^{242-262}$Fm with the available experimental data  \cite{Refs-barriers-other-nuclei-3-tsf}.
For the convenience of the reader, and also to facilitate the comparison with 
Uranium isotopes, we keep the style of our discussions as close as possible to 
the one used in Ref. \cite{Rayner-Robledo-fission-U}.

%
%
\begin{figure}
\includegraphics[width=0.47\textwidth]{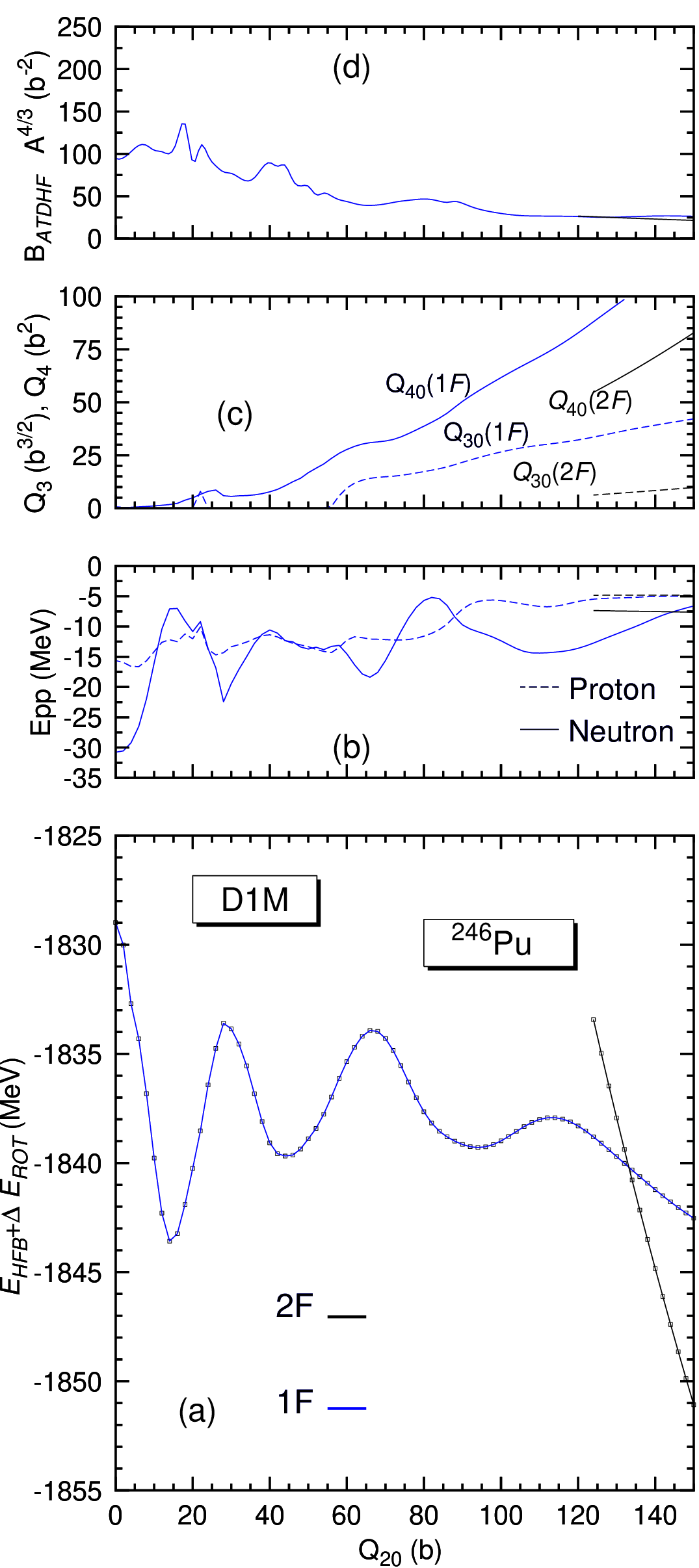}
\caption{ (Color online) The HFB plus the zero point rotational energies obtained 
with the Gogny-D1M EDF are plotted in panel (a) as functions of the quadrupole 
moment $Q_{20}$ for the nucleus $^{246}$Pu. Both the one (1F) and two-fragment (2F) solutions 
are included  in the plot. The pairing interaction energies are depicted 
in panel (b) for
protons (dashed lines) and neutrons (full lines). The octupole and hexadecapole moments 
corresponding to the 1F and 2F solutions are given in panel (c). The collective masses
obtained within the ATDHFB approximation are plotted in panel (d). For more
details, see the main text.
}
\label{example246Pu} 
\end{figure}

The paper is organized as follows. In Sec. \ref{Theory-used}, we briefly outline the theoretical 
framework used in the present study. For more details the interested reader is referred to 
Ref. \cite{Rayner-Robledo-fission-U}. In this section, we will also compare Gogny-D1M 
spontaneous fission half-lives for the nuclei $^{242-262}$Fm with the available 
experimental values. The results of our calculations for the isotopes 
$^{232-280}$Pu are discussed in Sec. \ref{RESULTS}. First, in Sec. \ref{Convergence}, we 
discuss the convergence of the calculations in terms of the basis size in the case
of the very neutron-rich nucleus $^{280}$Pu. We start section \ref{FB-systematcis}, with a 
detailed description of our fission calculations for the nucleus $^{246}$Pu taken as an 
illustrative example. Next, we present the systematics of the fission paths, 
spontaneous fission half-lives and fragment's charge and mass
 for the considered Plutonium isotopes.
In the same section we will also discuss the appearence of second isomeric states in  
Plutonium nuclei.
In Sec. \ref{change-pairing-strenght}, we explicitly discuss the impact of pairing correlations 
on the predicted  spontaneous fission half-lives for  $^{232-280}$Pu by increasing artificially the 
pairing strengths of the original Gogny-D1M EDF. Conclusions and work perspectives are presented 
in Sec. \ref{conclusions}.

\begin{figure*}
\includegraphics[width=0.3\textwidth]{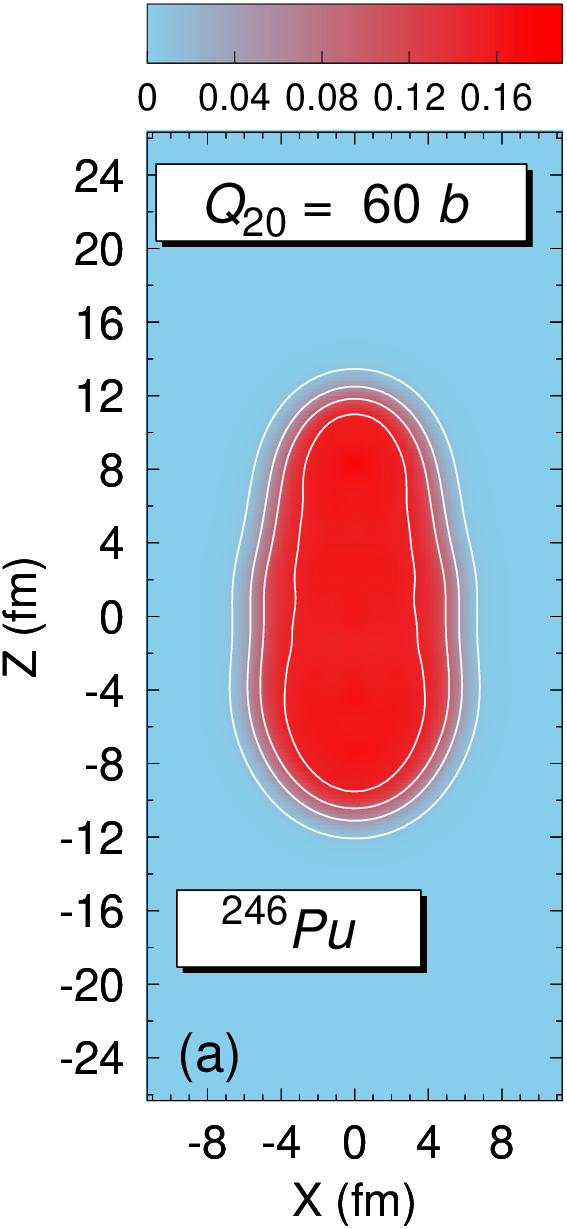} 
\includegraphics[width=0.3\textwidth]{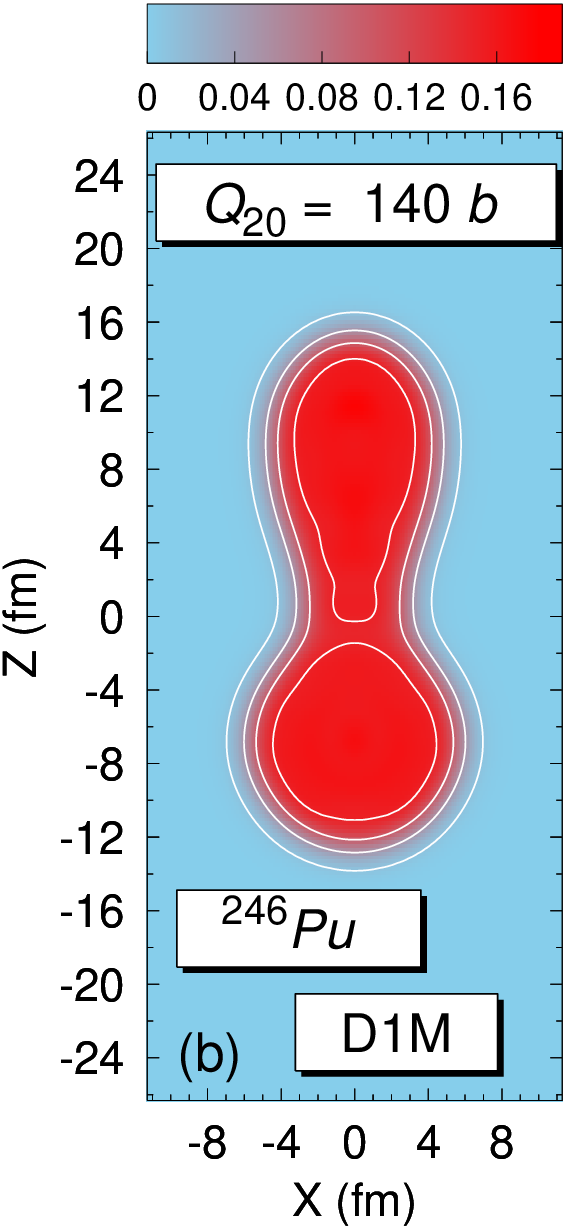} 
\includegraphics[width=0.3\textwidth]{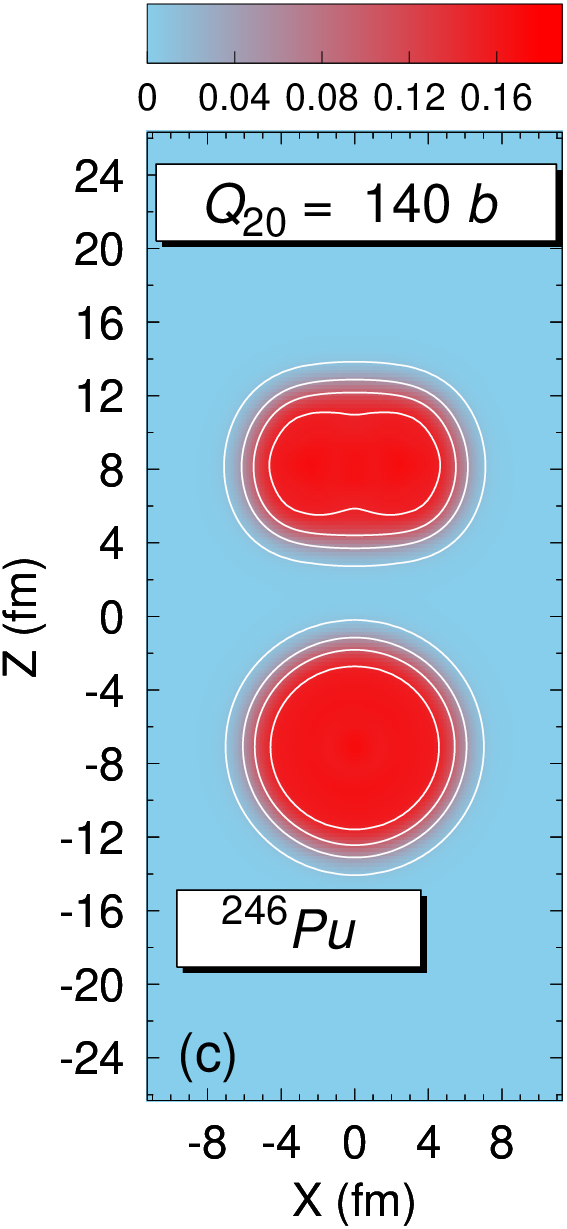} 
\caption{(Color online) Density contour plots for the isotope $^{246}$Pu at the quadrupole 
deformations Q$_{20}$=60 b [panel (a)] and Q$_{20}$=140 b [panels (b) and (c)]. The density 
contours in panels (a) and (b) correspond to 1F solutions of the HFB equations while 
the one in panel (c) corresponds to a 2F solution. Results have been obtained 
with the parametrization D1M of the Gogny-EDF. Densities are in units of fm$^{-3}$ 
and contour lines are drawn at 0.01, 0.05, 0.10 and 0.15 fm$^{-3}$.
}
\label{contourdensity246} 
\end{figure*}

\section{Theoretical framework}
\label{Theory-used}

As already mentioned, we have resorted to the constrained HFB approximation 
\cite{rs}.
We have  used as constraining 
operators the axially symmetric quadrupole  $\hat{Q}_{20}=z^{2}-\frac{1}{2}\left(x^{2}+y^{2} \right)$
and 
 octupole $\hat{Q}_{30}=z^{3} -\frac{3}{2}\left(x^{2}+y^{2} \right)z$ \cite{PRCQ2Q3-2012,Robledo-Rayner-JPG-2012} 
operators to obtain the corresponding one-fragment (1F) solutions. On  the 
other hand, constraints on the necking $\hat{Q}_{Neck}(z_{0},C_{0})=\exp \Big[-\left(z-z_{0} \right)^{2}/C_{0}^{2}  \Big] $ operator
are used to reach two-fragment (2F) solutions \cite{Robledo-Giulliani,Warda-Egido-Robledo-Pomorski-2002,Rayner-Robledo-fission-U}.
We have also considered a constraint on the operator 
$\hat{Q}_{10}=r P_{1} \left(cos(\theta) \right)$, to avoid 
spurious effects associated to the center 
of mass motion \cite{PRCQ2Q3-2012,Robledo-Rayner-JPG-2012}. Finally the typical HFB constraints
 on both the proton and neutron numbers \cite{rs} are considered.

The  quasiparticle operators   \cite{rs} have been expanded in 
a deformed
axially symmetric 
harmonic oscillator  (HO) basis 
containing  states with $J_{z}$ quantum numbers up to 35/2 and up to
26 quanta in the z direction. The basis quantum numbers are
restricted by the condition

\begin{eqnarray}
2n_{\perp} + |m| +\frac{1}{q} n_{z} \le M_{z,Max}
\end{eqnarray}
with $M_{z,Max}$=17 and q=1.5.
For
each of the fission configurations in  $^{232-280}$Pu 
and $^{242-262}$Fm, the 
HO lengths $b_{z}$ and $b_{\perp}$ have been optimized 
so as to minimize the total HFB energies. Both the 
choice of the basis size and the minimization of the 
energy with respect to the oscillator lengths 
$b_{z}$ and $b_{\perp}$
lead to well converged relative energies (see below). 
For the solution of the HFB equations, an 
approximate second order gradient method  \cite{Robledo-Bertsch2OGM,PRCQ2Q3-2012,PTpaper-Rayner,Robledo-Rayner-JPG-2012} has
 been used. The Coulomb exchange term has been considered in the Slater approximation 
 \cite{CoulombSlater,Marta-CE}
 while the spin-orbit contribution to the pairing field has been 
neglected. 
 
\begin{figure*}
\includegraphics[width=1.0\textwidth]{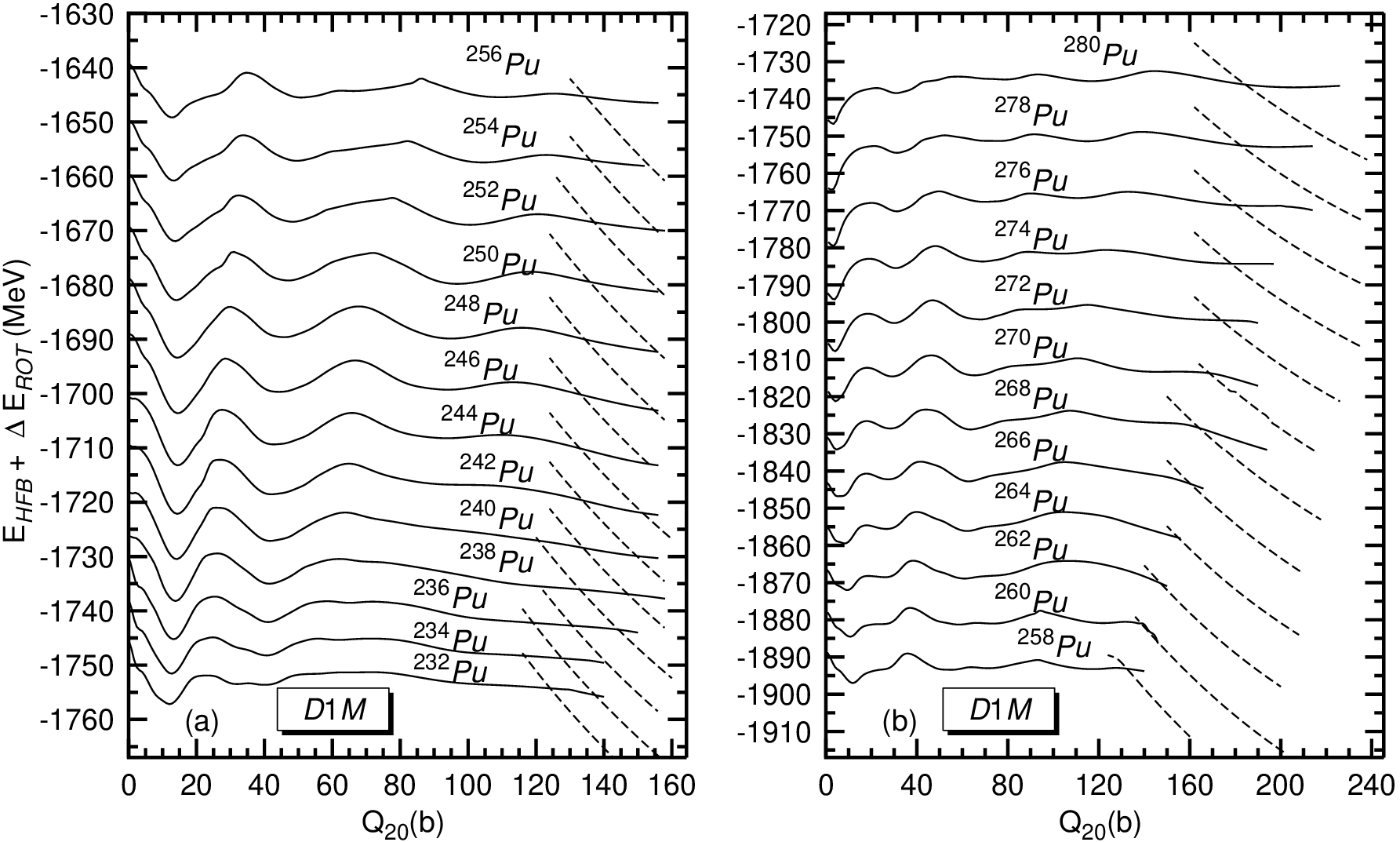} 
\caption{ The HFB plus the zero point rotational energies obtained 
with the Gogny-D1M EDF are plotted in panel (a) for the nuclei 
$^{232-256}$Pu and in panel (b) for
the nuclei
$^{258-280}$Pu
as functions of the quadrupole 
moment $Q_{20}$. Both the one (1F) and two-fragment (2F) solutions 
are shown  in the plot
with continuous and dashed lines, respectively. Starting from $^{234}$Pu ($^{260}$Pu) in
 panel (a) [in panel (b)] all
the curves have been successively shifted by 20 MeV in order to accomodate
them in a single plot. Note,  the different  energy scales  in both
panels.
For more
details, see the main text. 
}
\label{FissionBarriersD1M} 
\end{figure*}

 In order to obtain the corresponding fission
 paths for the considered Plutonium and Fermium nuclei, we have 
 employed the methodology outlined in our previous studies  
\cite{Robledo-Giulliani,Rayner-Robledo-fission-U}, i.e.,

\begin{itemize}

\item we have first carried out reflection-symmetric 
$Q_{20}$-constrained calculations. Subsequently, for each quadrupole deformation 
$Q_{20}$, we have constrained to a large  $Q_{30}$ value and then released such a constraint 
to reach the lowest energy solution. In this way, we have obtained 
the 1F solutions.

\item for sufficiently large quadrupole moments, we have  
constrained the number of
 particles in the neck of the parent nucleus to a small value and then released 
the constraint self-consistently. Calculations have been carried out  with different 
neck parameters $z_{0}$ and $C_{0}$ to ensure 
that the same lowest energy solution is always reached. In this way, we 
have obtained the 2F configurations for which the charge and  mass of the fragments 
lead to the minimum energy.

\end{itemize}

Though explicit 
 contraints are not included for them, the average values of 
 higher multipolarity moments (i.e., $\hat{Q}_{40}$, $\hat{Q}_{60}$, $\dots$ )
 are automatically adjusted
 during the self-consistent minimization of the HFB energy. 
 Note, that kinks and multiple 
branches are common in this type of calculations 
 \cite{Rayner-Robledo-fission-U,Dubray-discontinuities}
as a result of projecting 
multi-dimensional fission paths into a one-dimensional plot. 
Second, for the same 
reason, the 1F and 2F curves appear as intersecting ones. However,  in the multidimensional
 space of deformation parameters $\left(\hat{Q}_{20},\hat{Q}_{30},\hat{Q}_{Neck}, \dots \right)$, there 
 is a path with a ridge connecting them \cite{gogny-d1s}. We have 
 neglected the small  contribution of such a path to the action [see, Eq.(\ref{Action}) below]
 which amounts to
 take the 2F curves as really intersecting the 1F ones 
 \cite{Robledo-Giulliani,Rayner-Robledo-fission-U}.

\begin{figure}
\includegraphics[width=0.45\textwidth]{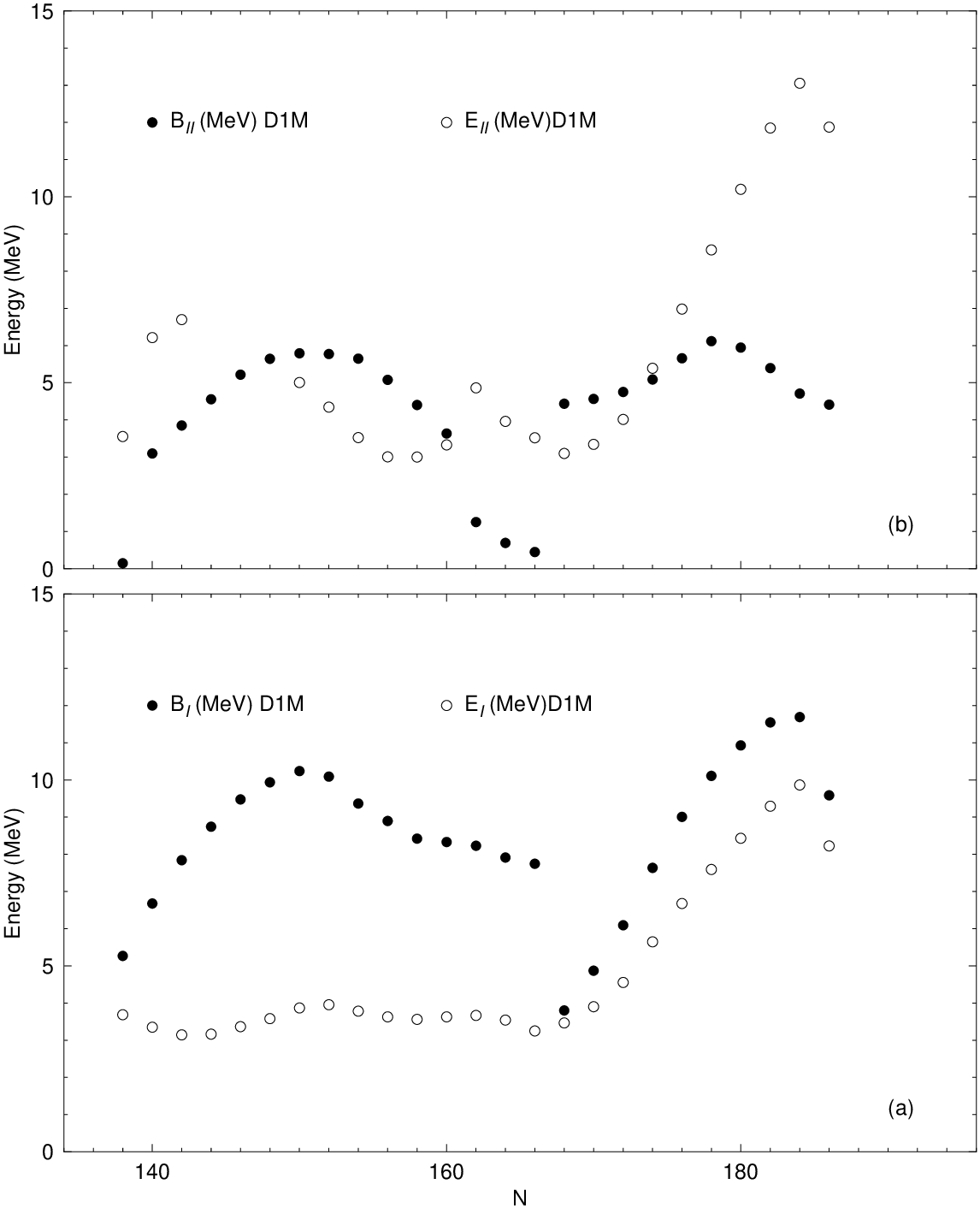} 
\caption{The excitation energies E$_{I}$ (E$_{II}$) and 
the barrier heights  B$_{I}$ (B$_{II}$) for the first
(second) isomeric wells in $^{232-280}$Pu are plotted 
in panel (a) [panel (b)]
as functions 
of the neutron number N.
}
\label{isomers-D1M} 
\end{figure}

The constrained  HFB calculations provide  all the 
ingredients required to obtain the collective masses 
and the zero point energy quantum corrections. 
To compute  the collective mass $B(Q_{20})$ and 
the zero point vibrational correction
$\Delta E_{vib}(Q_{20})$ 
the perturbative cranking approximation 
to both the Adiabatic Time Dependent HFB (ATDHFB) approach
\cite{crankingAPPROX,Giannoni,Libert-1999}  and the
Gaussian Overlap Approximation (GOA) to the GCM \cite{rs}
have been used.
The  rotational correction $\Delta E_{ROT}(Q_{20})$ has been 
expressed in terms of the Yoccoz moment of 
inertia \cite{RRG23S,ER-Lectures,NPA-2002}. For  details, the reader is 
referred to our previous work  \cite{Rayner-Robledo-fission-U}.

Within the standard Wentzel-Kramers-Brillouin (WKB)
formalism \cite{Baran-TSF-1,Baran-TSF-2}, the spontaneous fission half-life 
(in seconds) is given by

\begin{eqnarray} \label{TSF}
t_{SF}= 2.86 \times 10^{-21} \times \left(1+ e^{2S} \right)
\end{eqnarray}
where the action S along the quadrupole constrained fission path reads

\begin{eqnarray} \label{Action}
S= \int_{a}^{b} dQ_{20} \sqrt{2B(Q_{20})\left(V(Q_{20})-\left(E_{GS}+E_{0} \right)  \right)}
\end{eqnarray}  

The integration limits a and b correspond to the classical turning points
for the energy $E_{GS}+E_{0}$. The collective potential  $V(Q_{20})$ is given by 
the HFB energy 
corrected by the  
zero point rotational $\Delta E_{ROT}(Q_{20})$ and vibrational $\Delta E_{vib}(Q_{20})$ energies.

For the parameter $E_{0}$, we have considered four different  
values, i.e.,  $E_{0}$=0.5, 1.0, 1.5 and 2.0 MeV
\cite{Rayner-Robledo-fission-U}.
In order to analyze the impact of pairing correlations, on both  the zero point quantum fluctuations 
and the collective masses \cite{proportional-1,proportional-2}, we have also performed 
self-consistent
calculations for the isotopes $^{232-280}$Pu
with a modified Gogny-D1M EDF in which 
the pairing strengths have been increased by
5 $\%$  and 10 $\%$, by means of the 
same multiplicative factor ($\eta$=1.05 and 1.10, respectively)
in front of the proton and neutron pairing fields \cite{rs}.
As decay modes 
spontaneous fission and $\alpha$-decay compete and determine 
the stability of heavy nuclear systems \cite{WN-Nature,Erler2012,Warda-Egido-2012}. We
have computed the  $\alpha$-decay half-lives $t_{\alpha}$
using the  parametrization of the Seaborg-Viola formula given in Ref. \cite{TDong2005}.
The choice of D1M over D1S is specially justified here as a good description of 
$Q_{\alpha}$ values is essential for the Seaborg-Viola formula to perform well.

As already mentioned in the present study we have resorted to the parametrization D1M
of the Gogny-EDF  whose fitting protocol \cite{gogny-d1m} included both realistic 
neutron matter equation of state (EoS) information and the binding energies of all 
known nuclei. In this way, the Gogny-D1M EDF cures a known deficiency of the 
more standard 
D1S parametrization \cite{gogny-d1s}, i.e., a systematic drift in the differences 
between experimental and theoretical binding energies in heavy nuclei \cite{Hilare-2007}.
This is quite relevant if one keeps in mind that 
we will extrapolate to very neutron-rich Plutonium isotopes. This is the main reason 
underlying our choice of the Gogny-D1M EDF in the present study.

To further 
validate the use of the Gogny-D1M EDF, we have extended our previous t$_{SF}$ calculations 
(within the GCM and ATDHFB schemes)
for Fermium nuclei \cite{Rayner-Robledo-fission-U} to the  whole set of 
isotopes $^{242-262}$Fm  for which experimental data are available
\cite{Refs-barriers-other-nuclei-3-tsf}. This chain of isotopes has been studied 
in previous works. It is considered a very challenging testing ground 
with competing fission paths (see, for example, Refs. \cite{Warda-Egido-Robledo-Pomorski-2002,Warda-Egido-Robledo-Pomorski-2002,Fm-Doba}
and references therein). We have determined the 1F and 2F curves as well as all the required 
quantities along the lines described in this section.
 As can be seen from Fig.\ref{fig-Fm}, the predicted t$_{SF}$ values
nicely follow the bell-shaped experimental curve. These results corroborate our previous 
findings \cite{Rayner-Robledo-fission-U}, i.e., though uncertainties in the corresponding absolute values are large, the 
Gogny-D1M HFB framework captures  the behavior of fission 
observables like the spontaneous fission half-lives along isotopic chains
and represents a reasonable starting point to describe fission in heavy and superheavy nuclei. With this
in mind, we have carried out fission calculations for the isotopes 
$^{232-280}$Pu.

\section{Discussion of the results}
\label{RESULTS}
In this section, we present  the results of our 
Gogny-D1M
calculations. In 
Sec. \ref{Convergence}, we discuss the convergence 
in terms of the basis size in the case of the
very
neutron-rich
isotope $^{280}$Pu. First, in Sec. \ref{FB-systematcis}, we discuss in detail our results for 
the nucleus $^{246}$Pu, taken as an illustrative example. Subsequenly, we
present the systematics  of our fission calculations for $^{232-280}$Pu. 
Finally, in Sec. \ref{change-pairing-strenght}, we will discuss the impact of pairing correlations 
on the predicted  $t_{SF}$ values  using a modified Gogny-D1M EDF.

\subsection{Convergence of the calculations}
\label{Convergence}

In our calculations, bases with $M_{z,MAX}$=13, 14, 15, 16, 17 and 18 have been used
to check the convergence of the results. In all cases we have considered the value q=1.5
and optimized the HO lengths $b_{z}$ and $b_{\perp}$. In Fig. \ref{fig-convergence} (a) 
 we have plotted the rotationally corrected energies
E$_{HFB}$ + $\Delta$ E$_{ROT}$ corresponding to the 1F configurations in 
$^{280}$Pu as functions of the quadrupole moment $Q_{20}$.
The vibrational energy corrections 
$\Delta$ E$_{vib}$
have not been included in the plot
since they are rather constant as functions of the quadrupole moment.
The inset in panel (a) displays, for each $M_{z,MAX}$, the relative energies 
referred to the corresponding ground states. On the other hand, Fig. \ref{fig-convergence} (b) depicts 
the energy differences with respect to the calculations with  $M_{z,MAX}$=18.

\begin{figure}
\includegraphics[width=0.450\textwidth]{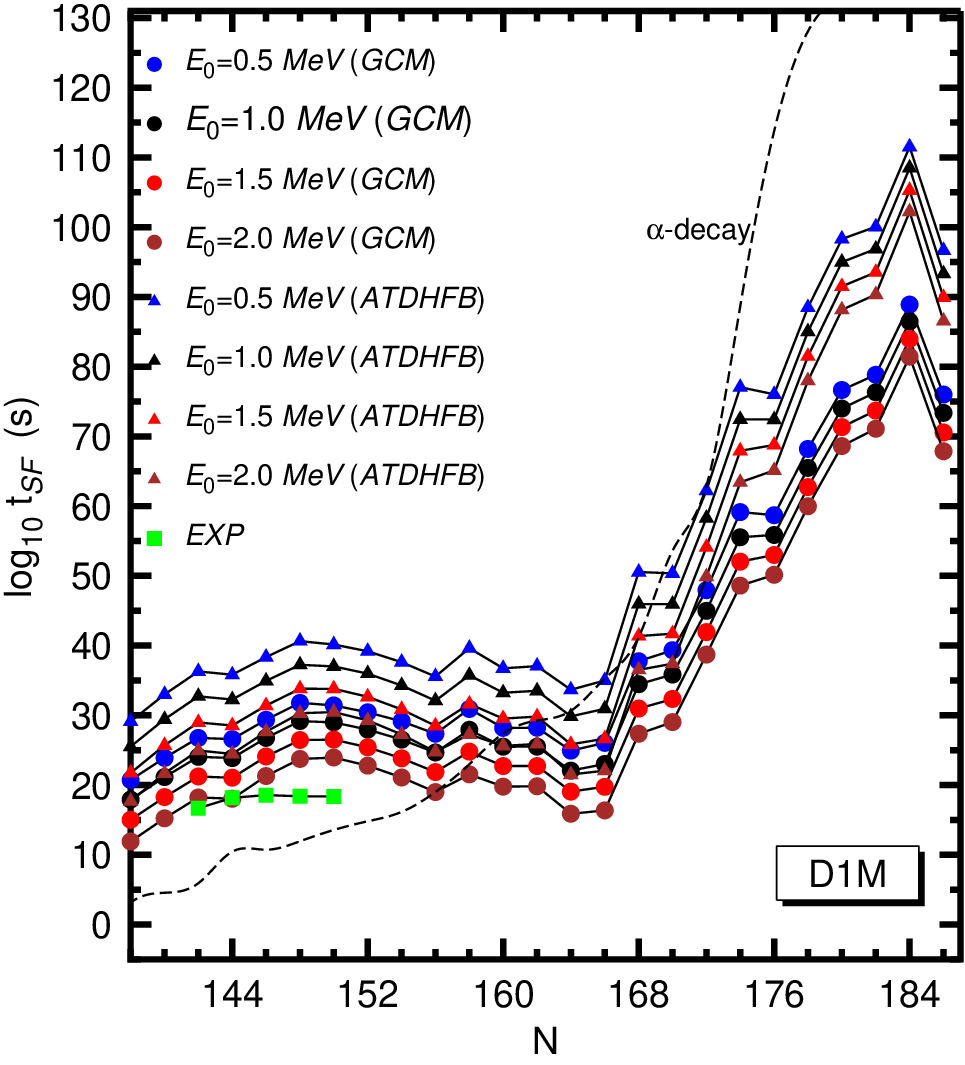} 
\caption{(Color online) The spontaneous fission half-lives $t_{SF}$, 
predicted within the GCM and ATDHFB schemes, 
for the isotopes $^{232-280}$Pu  are 
 depicted as functions of the neutron number. Results have been obtained with 
 the  Gogny-D1M EDF. Calculations have been carried out  with 
 $E_{0}$=0.5, 1.0, 1.5 and 
 2.0 MeV, respectively. The experimental $t_{SF}$ values 
 \cite{Refs-barriers-other-nuclei-3-tsf}
 for  $^{236-244}$Pu
 are included in the plot. In addition,  $\alpha$-decay half-lives are plotted
 with  short dashed lines. For more details, see the main text.
}
\label{tsf-D1M} 
\end{figure}

From Fig.\ref{fig-convergence} (a) one concludes that the bases with $M_{z,MAX}$=13, 14 and  15
are too small to describe the 1F configurations in this 
very neutron-rich  isotope at very large quadrupole 
deformations. 
On the other hand, larger bases with $M_{z,MAX}$=16, 17 and 18 
already provide quite similar profiles for the 1F curves. This is further 
corroborated from the relative energies shown in the inset. Note, that such 
relative energies   
are the ones determining the dynamics of the fission process instead of their 
absolute values. In fact, a small basis with $M_{z,MAX}$=13 is enough 
to accurately describe 1F configurations  up to $Q_{20}$ $\approx$ 80 b
while for larger $Q_{20}$ values convergence is only achieved by 
increasing the basis size. The energy differences $\delta \epsilon$ 
in Fig.\ref{fig-convergence} (b), show that even for very large $Q_{20}$ values around 200 b, the
basis with $M_{z,MAX}$=17 provides an error (with respect to $M_{z,MAX}$=18) 
always smaller than 0.81 MeV.
Similar or even more accurate results also hold for lighter Plutonium isotopes.
We have therefore used  a basis with $M_{z,MAX}$=17 \cite{Rayner-Robledo-fission-U}
in all the calculations discussed in 
the following sections as it  provides a reasonable compromise between 
accuracy and the computational effort required to describe 
the  fission paths in $^{232-280}$Pu.

\subsection{Systematics of fission paths, spontaneous fission half-lives and 
fragment mass 
in Plutonium isotopes}
\label{FB-systematcis}

In this section, we discuss  the systematics of our calculations for
the isotopes $^{232-280}$Pu. Let us first describe 
in more detail
the results 
obtained for the nucleus $^{246}$Pu, taken as an illustrative
example. 
In Fig. \ref{example246Pu} (a), we show the energies 
 E$_{HFB}$ + $\Delta$ E$_{ROT}$, as functions of
  Q$_{20}$, for the 1F and 2F solutions, respectively. The ground state is located at Q$_{20}$=14 b while 
a first fission isomer appears at Q$_{20}$=44 b with an excitation energy
of 3.90 MeV. This first isomer is separated from the  ground state by an inner barrier, the top of which is located around Q$_{20}$=28 b, whose  
height amounts to 9.98 MeV. In our previous study 
\cite{Rayner-Robledo-fission-U}, we have already explored the well kown
reduction of the inner barrier due to triaxiality \cite{Abusara-2010,Delaroche-2006}
for a selected set of Uranium, Plutonium and  superheavy nuclei  for 
which experimental data are available 
\cite{Refs-barriers-other-nuclei-1,Refs-barriers-other-nuclei-2,Refs-barriers-other-nuclei-3-tsf,Pu-mass-fragments-exp-1,Pu-mass-fragments-exp-2}. Such a 
lowering of the inner barriers comes with an increase of the 
collective inertia  \cite{Baran-1981,Bender-1998}
 that tends to compensate the value of the action.
As a result, the influence of triaxiality on the predicted 
spontaneous fission half-lives is quite limited 
\cite{WN-Nature,Baran-1981}
and has not been 
considered in the present study. From Fig. \ref{example246Pu} (a), one also observes the second and third 
fission barriers as well as a second fission isomer in between them
at Q$_{20}$=94 b. This second isomer lies 4.29 MeV above the 
ground state. As we will see later on, second fission isomers 
are also obtained for other Plutonium nuclei \cite{Rayner-Robledo-fission-U}.

\begin{figure}
\includegraphics[width=0.450\textwidth]{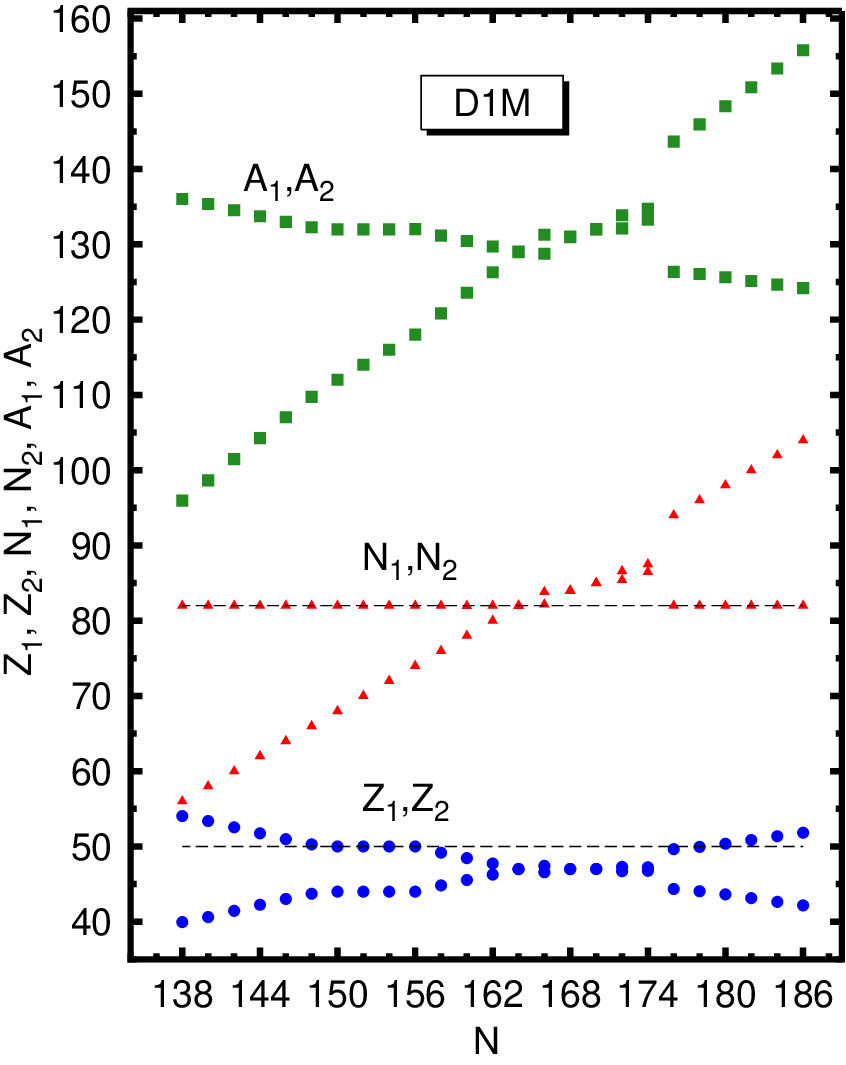} 
\caption{(Color online) The proton ($Z_{1},Z_{2}$), neutron ($N_{1},N_{2}$) and mass ($A_{1},A_{2}$)
numbers of the two  fragments resulting from the fission of the isotopes $^{232-280}$Pu 
are shown as functions of the neutron number in the  parent nucleus.
Results have been obtained with the Gogny-D1M 
EDF. The magic proton Z=50 and  neutron N=82 numbers
are highlighted with dashed horizontal lines to guide the eye.  
}
\label{mass-dist-D1M} 
\end{figure}

The proton (dashed lines) and neutron (full lines) pairing interaction energies \cite{rs} are shown in 
Fig. \ref{example246Pu} (b). The neutron energies exhibit 
minima at the spherical configuration, around the top of the inner and 
second fission barriers as well as around Q$_{20}$=110 b. On the other hand, the values
of the 
octupole and hexadecupole moments corresponding,
 to the 1F 
[i.e., Q$_{30}(1F)$ and Q$_{40}(1F)$]  and 2F [i.e., Q$_{30}(2F)$ and Q$_{40}(2F)$] curves
in
Fig. \ref{example246Pu} (c)
clearly reflect the separation of those paths in the multidimensional
space of parameters.

In Fig. \ref{example246Pu} (d), we have plotted the 
collective masses obtained within the ATDHFB scheme. 
The GCM masses (not shown in the figure) display a similar trend 
but are, on the average, always smaller than the corresponding ATDHFB values.  
Such differences between the ATDHFB and GCM masses  have also been found 
in previous studies \cite{Robledo-Giulliani,Rayner-Robledo-fission-U,Baran-TSF-2}
and can lead to differences of several orders of magnitud in the 
predicted t$_{SF}$ values. This is the reason why both the ATDHFB and GCM 
collective
masses have been used in the present work to compute spontaneous 
fission half-lives. One should also keep in mind, that the collective 
inertias are computed in the perturbative cranking scheme
\cite{Rayner-Robledo-fission-U,crankingAPPROX,Giannoni,Libert-1999}. 
For example, for E$_{0}$=1.5 MeV, the t$_{SF}$ values predicted
within the ATDHFB and GCM schemes are 4.544 $\times$ 10$^{32}$ s
and 2.581 $\times$ 10$^{25}$ s, respectively. 
Let us also mention, that the wiggles in the masses have been softened 
using a three point filter \cite{Rayner-Robledo-fission-U}. 
Since we take the 
1F and 2F curves as intersecting and do not include the effect 
of the $\gamma$ degree of freedom,  the  t$_{SF}$ 
values reported in this work should be taken 
as lower bounds to the real ones. 

In Fig. \ref{contourdensity246}, we have plotted the density profiles
for the nucleus $^{246}$Pu at the 1F configurations with Q$_{20}$=60 
and 140 b  [panels (a) and (b)]. On the other hand, the corresponding 
2F solution at Q$_{20}$=140 b is shown in Fig. \ref{contourdensity246} (c). It
consists of a spherical $^{132}$Sn fragment and an oblate and slightly
octupole deformed $^{114}$Ru fragment with $\beta_{2}$=-0.23 and 
$\beta_{3}$=0.02 (referred to the fragment's center of mass). The
oblate shape of the $^{114}$Ru fragment minimizes a large
Coulomb repulsion of 205.81 MeV.

%
%
\begin{figure*}
\includegraphics[width=0.32\textwidth]{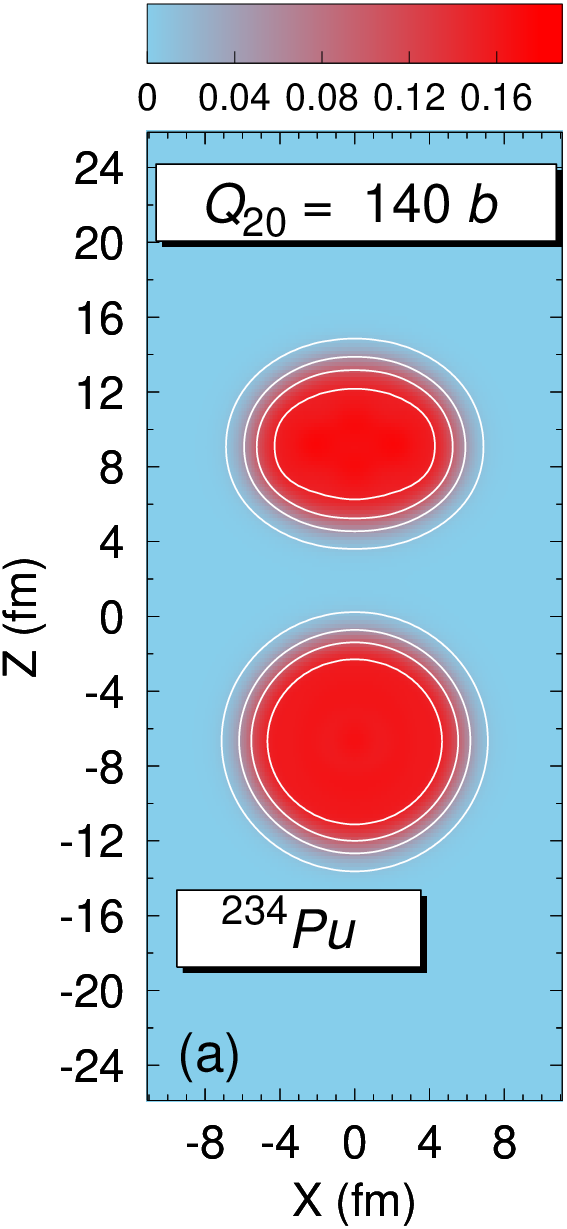}
\includegraphics[width=0.32\textwidth]{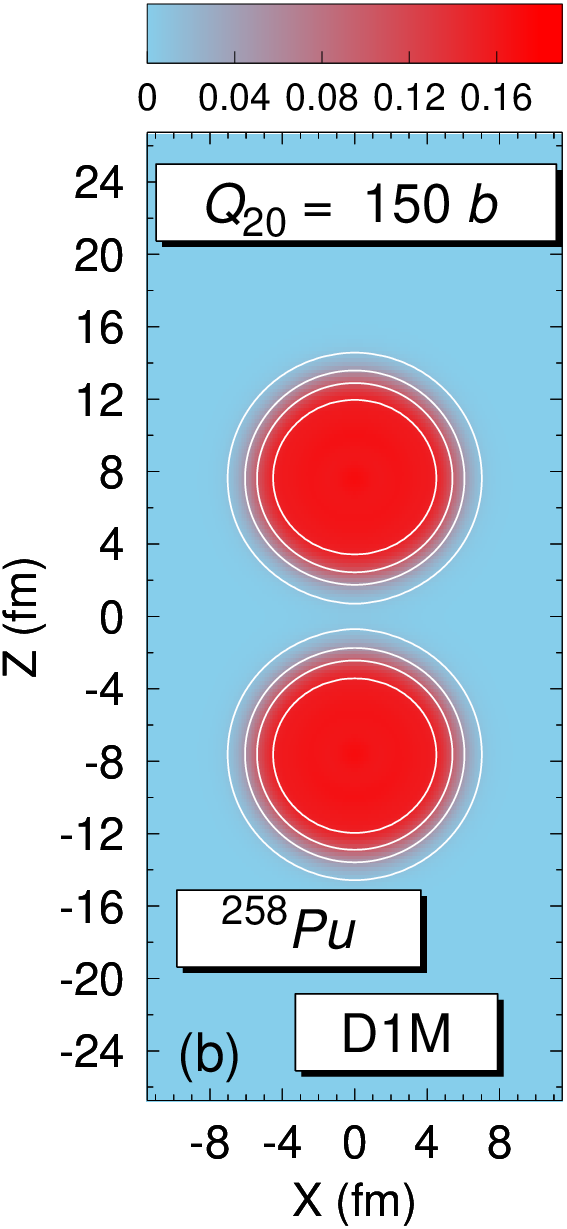}
\includegraphics[width=0.32\textwidth]{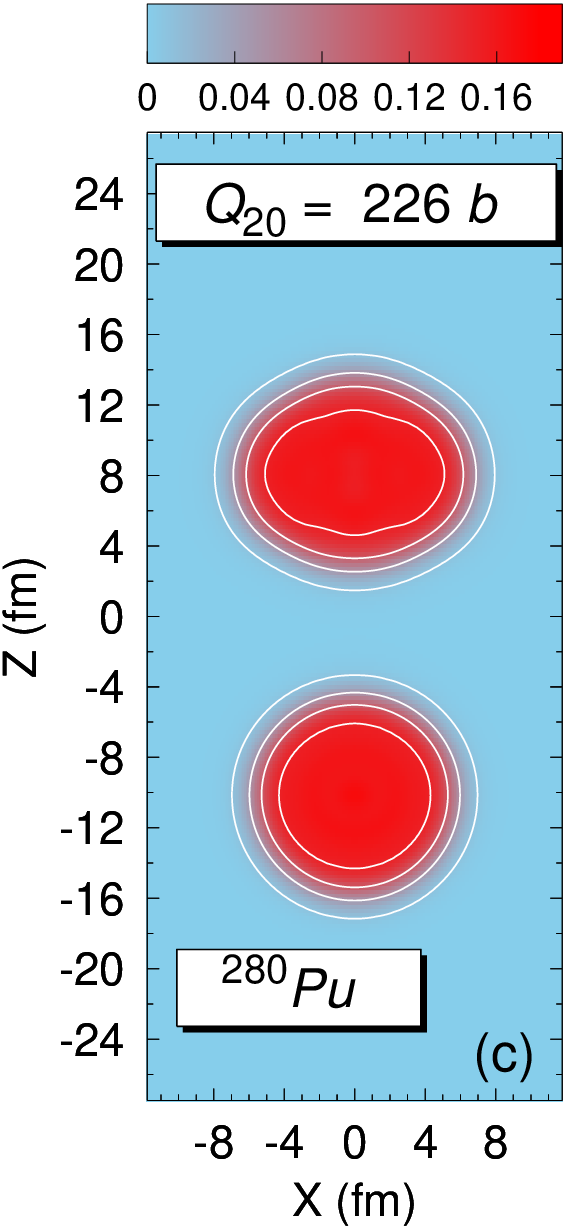}
\caption{ (Color online) Density contour plots for the nuclei
 $^{234}$Pu [panel (a)],  $^{258}$Pu [panel (b)] an  $^{280}$Pu [panel (c)].
 The density profiles correspond to 2F solutions at the quadrupole deformations 
 $Q_{20}$=140, 150 and 226 b, respectively.
 Results are shown 
for the Gogny-D1M EDF. Densities are in units of fm$^{-3}$ 
and contour lines are drawn at 0.01, 0.05, 0.10 and 0.15 fm$^{-3}$. 
}
\label{den_cont_PU_D1M} 
\end{figure*}

Some comments are in order here. First, as we will see later on, oblate deformed fragments 
are also obtained in our calculations for other Plutonium isotopes. 
Similar results have already been obtained in previous studies of the 
 Uranium 
isotopes \cite{Robledo-Giulliani,Rayner-Robledo-fission-U}  and deserve 
further attention as only prolate deformations are usually assumed 
for fission fragments
\cite{Moller-1,Moller-2}. Second, as discussed in previous works
\cite{Nenoff-2007,Piessens-1993,Ter-1996}, the likelihood 
of obtaining the 
$^{132}$Sn fragment is related to the 
key role played by the magic 
proton Z=50 and neutron N=82 numbers. This is 
not surprising as such a distribution is obtained applying the 
Ritz variational principle \cite{Blaizot-Ripka} to the corresponding 
HFB energy. However, the comparison with the experimental data 
\cite{Pu-mass-fragments-exp-1,Pu-mass-fragments-exp-2} 
reveals
that the
2F configurations resulting from minimizing 
the HFB energy are not necessarily the ones arising after
scission. For example, for nuclei
in the considered region of the nuclear chart, the experimental
mass number of the heavy fragment is close to A=140 instead of the 
value A=132 obtained in our calculations.
In our (minimal energy) calculations the properties of the 
fragments are determined from 2F solutions at the largest 
quadrupole moments.
If, on the other hand, we take the breaking point as the one 
where the neck reaches a critical value  \cite{Chasman-breaking}
the predicted heavy fragment mass number  turns out to be closer 
to the experimental one \cite{Rayner-Robledo-fission-U}. Last, but not least, the predicted masses 
should be taken as an approximation to the peaks of the experimental broad mass
distribution of the fragments. In order to account for 
prescission quantum shell effects as well as the 
broad mass  distribution a more sophisticated (dynamical) approach than ours is 
needed (see, for example, \cite{Goutte-dynamical-distribution,Goriely-dist} and 
references therein). We will not pursue this kind of computationally 
involved approximation in the present study and simply keep in mind 
that the mass distribution of the fragments leading to the minimal 
HFB energy slightly underestimates the heavy fragment mass.

In Fig. \ref{FissionBarriersD1M} 
we have plotted the energies E$_{HFB}$ + $\Delta$ E$_{ROT}$ for the nuclei $^{232-256}$Pu  [panel (a)] and 
$^{258-280}$Pu [panel (b)]. Both the  1F (full lines)  and 2F (dashed lines) curves  
are shown in the plots. Starting from $^{234}$Pu ($^{260}$Pu) in panel (a) [in panel (b)] all
the curves have been successively shifted by 20 MeV in order to 
accomodate them in a single plot. The first apparent feature from the figure, is the 
gradual
decreasing 
of the 
ground state deformations 
as we move towards the neutron dripline reaching Q$_{20}$=2-6 b in the heavier 
isotopes. Note, that the deformed ground states in 
$^{272-280}$Pu are a direct consequence of the approximate 
restoration of the broken rotational symmetry \cite{Rayner-Robledo-fission-U,RRG23S,NPA-2002,Rayner-PRC2004}
.
 However, from the (intrinsic)
HFB point of view the nuclei $^{272-280}$Pu  are spherical. The 
neutron pairing energy only vanishes at Q$_{20}$=0 for $^{278}$Pu. In addition, 
we have 
computed the two-neutron separation energies that reveal a sudden drop at N=186. Both 
are clear signatures of the magicity of the neutron number N=184.  On the 
 other hand, the inner barrier heights increase and the 1F curves widen 
 for increasing neutron number. 
The previous results 
agree well with the ones obtained for Uranium isotopes 
\cite{Robledo-Giulliani,Rayner-Robledo-fission-U} as well as with
the Extended Thomas-Fermi calculations of Ref. \cite{Tomas-Fermi}
which predicted
very high
barrier heights for N=184 isotones in this 
region of the nuclear chart.

Another prominent feature from  Fig.\ref{FissionBarriersD1M} is the appearence 
of second fission isomers   in the 
1F curves 
of several of the considered Plutonium isotopes. Such second isomers have  been
predicted within the microscopic-macroscopic (MM) approach 
\cite{Cwiok-third-min,Ben-third-min,Pask,Moller-Nucl-Phys-1972,Kowal-thir-min}
as well as
in several self-consistent calculations \cite{Delaroche-2006,Berger-thir-min,Rutz-thir-min}. They
have also been found in our previous HFB study \cite{Rayner-Robledo-fission-U} for the nuclei $^{232-280}$U
regardless of 
the particular version 
of the Gogny-EDF employed. Moreover, the results discussed in the present work and the 
ones in Refs. \cite{Robledo-Giulliani,Rayner-Robledo-fission-U,Mcdonell-2}, based on different 
 EDFs, show that 
the shell effects leading to fission isomers in 
the corresponding 1F curves of Uranium, Plutonium and Thorium nuclei   
are systematically present in different mean-field calculations. The issue
of why mean-field calculations do not reproduce the scarce experimental
data deserves further consideration.

In Fig. \ref{isomers-D1M}, we have plotted 
the excitation energies E$_{I}$ (E$_{II}$) and 
the barrier heights  B$_{I}$ (B$_{II}$) for the first
(second) isomeric wells 
in panel (a) [panel (b)], as functions 
of the neutron number N, for the nuclei $^{232-280}$Pu. From Fig. \ref{isomers-D1M} (a), we 
observe that the barrier heights B$_{I}$
exhibit a sudden drop at N=168 while the
excitation energies
E$_{I}$ of the first fission isomers remain relatively constant 
up to the same neutron number. For larger neutron numbers E$_{I}$ increases linearly up
to N=184 where both E$_{I}$ and B$_{I}$ display a sudden drop which 
is characteristic of the filling of a new major shell.  The 
barrier heights B$_{II}$, shown in Fig. \ref{isomers-D1M} (b), display two maxima, one 
at N=150 and the other at N=178. On the other hand, similar
to E$_{I}$, the excitation energies E$_{II}$ increase linearly for 
N $\ge$ 168
and display a sudden drop at N=184. Another relevant feature from Fig. \ref{isomers-D1M} (b) is the lack of
a second isomeric well for some light isotopes. For a comparison of the excitation energies of fission
isomers in $^{238-242}$Pu, the reader is referred to Ref. \cite{Rayner-Robledo-fission-U}.

In our previous 
work \cite{Rayner-Robledo-fission-U}, we have also explored the role 
of the $\gamma$ degree of freedom for configurations around the top
of the inner barrier in a selected set of nuclei for which experimental
data are available
\cite{Refs-barriers-other-nuclei-1,Refs-barriers-other-nuclei-2,Refs-barriers-other-nuclei-3-tsf,Pu-mass-fragments-exp-1,Pu-mass-fragments-exp-2}
. In the case of $^{238-244}$Pu, for example, triaxiality reduces the predicted
B$_{I}$ values by 1.11, 1.75, 2.23 and 2.74 MeV, respectively, though the theoretical 
values are still larger than the experimental ones. The same overestimation is also
observed for the 
 outer barriers, though the inclusion of reflection asymmetric shapes 
leads to a reduction of a few MeV. Ours and previous calculations for nuclei in this 
region \cite{Rayner-Robledo-fission-U,Delaroche-2006}, seem to suggest
that other effects not explicitly taken into account in this work
may be required to
improve the agreement with the available experimental data. Among them, the
pairing 
degrees of freedom and/or the collective dynamics appear as plausible candidates 
to be considered in future work. However, one should keep in mind that
the experimental data for barrier heights are model dependent
and therefore less reliable than the corresponding fission half-lives for a 
comparison with theoretical values.

Let us now turn our attention to the spontaneous fission half-lives 
predicted  for the isotopes $^{232-280}$Pu within the GCM and ATDHFB schemes.
They are depicted in Fig.\ref{tsf-D1M} 
 as functions of the neutron number. Calculations have been carried 
 out  with 
 $E_{0}$=0.5, 1.0, 1.5 and 
 2.0 MeV, respectively. The experimental $t_{SF}$ values 
 \cite{Refs-barriers-other-nuclei-3-tsf}
 for  $^{236-244}$Pu
 are also included in the plot.
The ATDHFB t$_{SF}$ values are always larger than the GCM ones. For example, 
for  $^{234}$Pu  (E$_{0}$=1.5 MeV) the  
GCM and ATDHFB values  are 
1.841 $\times$ 10$^{18}$ s and 4.864 $\times$ 10$^{25}$ s
while for $^{250}$Pu the corresponding values are
 7.384 $\times$ 10$^{21}$ s and 
3.146 $\times$ 10$^{28}$ s, 
respectively. The differences 
between the GCM and ATDHFB fission half-lives increase with increasing neutron number
reaching 22  orders of magnitud for $^{278}$Pu. On the other hand, increasing 
E$_{0}$ always leads to smaller t$_{SF}$ values. For the isotopes with neutron number
N $\ge$ 166 we observe a steady increase in the spontaneous fission half-lives 
reaching a maximum for the magic neutron number N=184.

%
%
\begin{figure}
\includegraphics[width=0.47\textwidth]{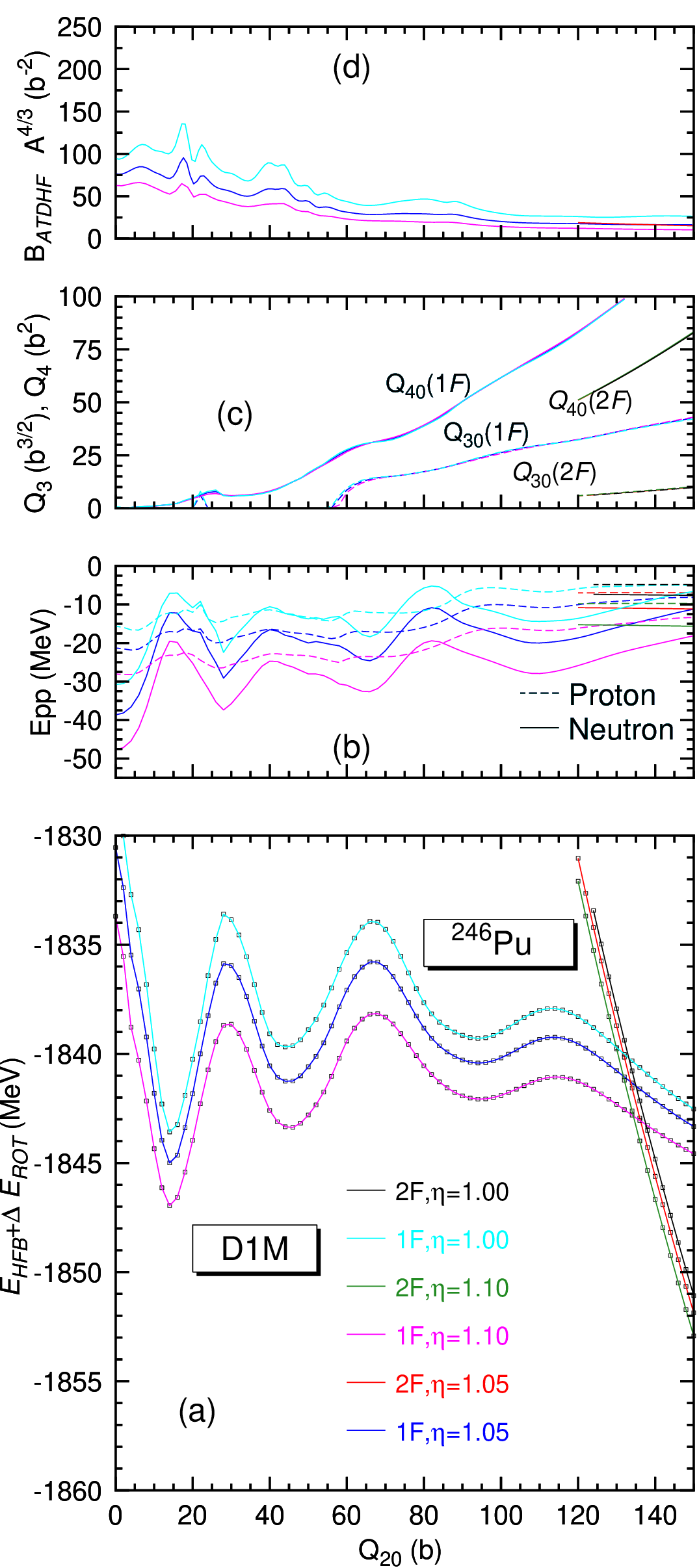}
\caption{ (Color online) The HFB plus the zero point rotational energies obtained 
with the normal ($\eta$=1.00) and modified ($\eta$=1.05 and 1.10)
Gogny-D1M EDFs are plotted in panel (a) as 
functions of the quadrupole 
moment $Q_{20}$ for the nucleus $^{246}$Pu. For each 
$\eta$ value, both the one (1F) and two-fragment (2F) solutions 
are included  in the plot. The pairing interaction energies are depicted 
in panel (b) for
protons (dashed lines) and neutrons (full lines). The octupole and hexadecapole moments 
corresponding to the 1F and 2F solutions are given in panel (c). The collective masses
obtained within the ATDHFB approximation are plotted in panel (d). For more
details, see the main text.
}
\label{FissionBarriers-eta-246Pu} 
\end{figure}

In Fig. \ref{tsf-D1M}, we have also plotted the $\alpha$-decay half-lives 
$t_{\alpha}$ computed with the  parametrization given in Ref. \cite{TDong2005}.
We have used the binding energies obtained for the corresponding Plutonium 
and Uranium nuclei. Let us stress that the parametrization D1M is well 
suited for such calculations since it has been tailored to provide a better
description of the nuclear masses 
\cite{gogny-d1m}
than the standard Gogny-D1S \cite{gogny-d1s} EDF.  
As can be seen, for increasing neutron number fission turns out to be 
faster than $\alpha$-decay. For Plutonium isotopes, our calculations 
predict the crossing point to be N $\approx$ 160, i.e., around two mass
units later than the D1M value found for Uranium isotopes   \cite{Rayner-Robledo-fission-U}.

\begin{figure*}
\includegraphics[width=1.0\textwidth]{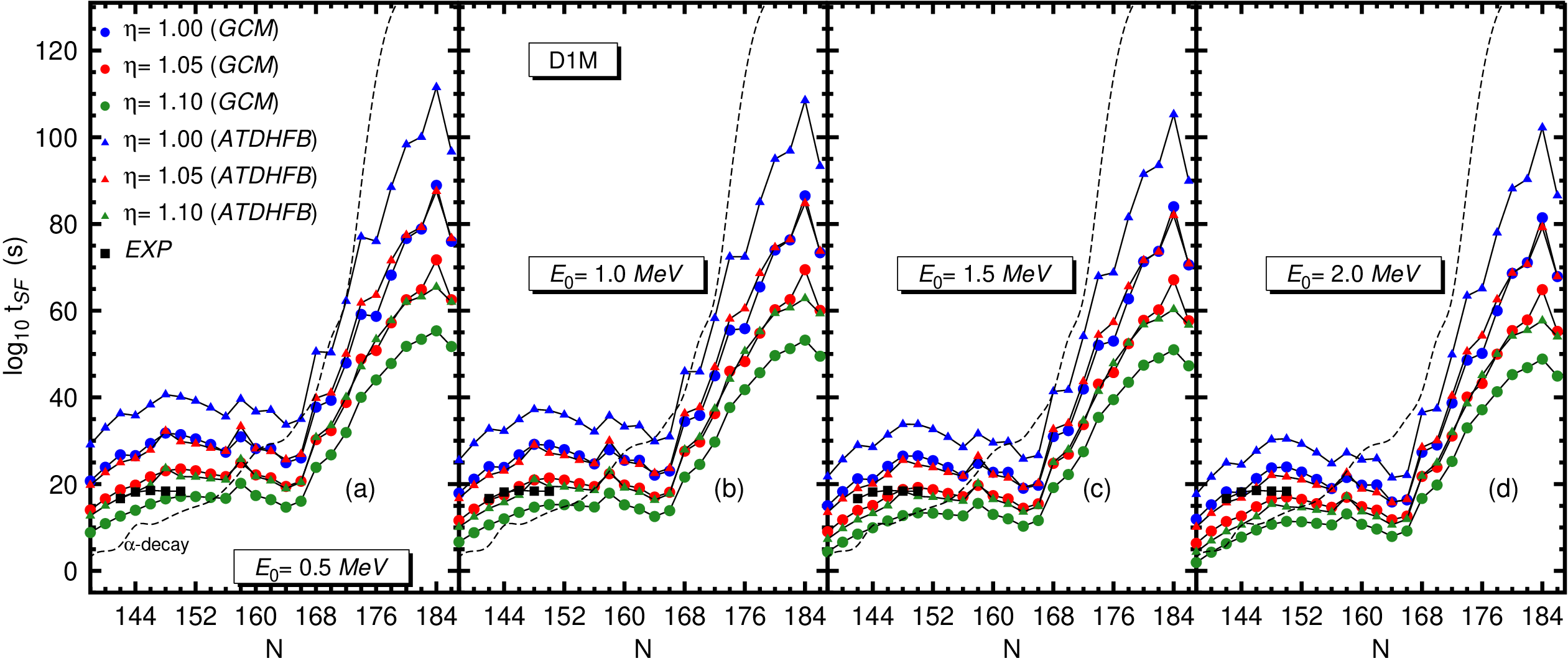} 
\caption{(Color online) The spontaneous fission half-lives $t_{SF}$, predicted within the GCM and ATDHFB schemes, 
for the isotopes $^{232-280}$Pu  are 
 depicted as functions of the neutron number. Results have been obtained with the
normal ($\eta$=1.00) and modified ($\eta$=1.05 and 1.10)
Gogny-D1M EDFs. Calculations have been carried out  with 
 $E_{0}$=0.5 [panel (a)], 1.0 [panel (b)], 1.5 [panel (c)] and 
 2.0 MeV [panel (d)], respectively. The experimental $t_{SF}$ values 
 \cite{Refs-barriers-other-nuclei-3-tsf}
 for  $^{236-244}$Pu
 are included in the plot. In addition,  $\alpha$-decay half-lives are plotted
 with  short dashed lines. For more details, see the main text.
}
\label{tsf-eta-D1M} 
\end{figure*}

In Fig.\ref{mass-dist-D1M} we have plotted the proton ($Z_{1},Z_{2}$), neutron
($N_{1},N_{2}$) and mass ($A_{1},A_{2}$) numbers corresponding to the 2F configurations 
in  $^{232-280}$Pu. In all cases, the 2F solutions have been taken for the largest
quadrupole deformations available so as to guarantee that fragment properties 
are nearly independent of the quadrupole moment. Once more, one clearly sees 
the key role played by both the neutron N=82 and proton Z=50 magic numbers
in the masses and charges  of the predicted  fission fragments. However, as already explained above in the case of 
$^{246}$Pu, in our calculations the properties of the fragments have been determined  using minimal energy
criteria. Therefore, caution should be taken when comparing with the experiment
\cite{Pu-mass-fragments-exp-1,Pu-mass-fragments-exp-2,Schmidt}.

Finally, in Fig. \ref{den_cont_PU_D1M} we have plotted the density contours 
for the nuclei $^{234}$Pu [panel (a)], $^{258}$Pu [panel (b)] and $^{280}$Pu [panel (c)].
The 2F solutions correspond to the quadrupole deformations Q$_{20}$=140, 150 
and 226 b, respectively. The lighter and heavier fragments in $^{234}$Pu
and $^{280}$Pu are predicted to be oblate ($\beta_{2}$=-0.21) and slightly octupole
($\beta_{3}$=0.01)
 deformed. Oblate deformed fragments have also been obtained for other 
 Plutonium and Uranium nuclei
 \cite{Robledo-Giulliani,Rayner-Robledo-fission-U}. They deserve further 
 study as only prolate deformations are usually assumed
 \cite{Moller-1,Moller-2}
  for fission fragments.
On the other hand, for $^{258}$Pu our Gogny-D1M calculations
predict a symmetric splitting into two spherical fragments.

\subsection{Varying pairing strengths 
in Plutonium isotopes}
\label{change-pairing-strenght}

In this section we explicitly consider the impact of pairing correlations on the predicted 
t$_{SF}$ values for $^{232-280}$Pu. To this end, both the proton and 
neutron pairing fields 
\cite{rs}
of the Gogny-D1M EDF ($\eta$=1) has been scaled
by the same factors $\eta$=1.05 and 1.10, respectively \cite{Robledo-Giulliani,Rayner-Robledo-fission-U}.

Let us briefly summarize our findings in the case of $^{246}$Pu, taken
as an illustrative outcome of our calculations. The rotationally corrected 1F and 2F HFB energies obtained with the 
normal ($\eta$=1.00) and modified ($\eta$=1.05 and 1.10)
Gogny-D1M EDFs are plotted in  
Fig. \ref{FissionBarriers-eta-246Pu} (a), as 
functions of the quadrupole 
moment $Q_{20}$. Regardless
of the $\eta$ value, the 1F and 2F curves display similar profiles  
in all the considered isotopes and are shifted downward 
with increasing $\eta$ values. As expected, the proton (dashed lines)
and neutrons (full lines), pairing interaction energies shown in 
Fig. \ref{FissionBarriers-eta-246Pu} (b), become larger with increasing 
$\eta$ values. On the other hand, the multipole moments shown in 
Fig. \ref{FissionBarriers-eta-246Pu} (c) are nearly independent of $\eta$
and lie on top of each other.

As a consequence of the inverse dependence of the collective masses with 
the square of the pairing gap \cite{proportional-1,proportional-2}, the 
ATDHFB collective masses, depicted in  
Fig. \ref{FissionBarriers-eta-246Pu} (d), are strongly correlated with 
the corresponding $\eta$ values. The same is also true for the GCM masses
(not shown in the plot). For example, the ATDHFB and GCM masses
are reduced by 30 $\%$ and 24 $\%$ for $\eta$=1.05 while 
for $\eta$=1.10 they are reduced by 50 $\%$ and 40 $\%$, respectively.
These reductions change the predicted spontaneous fission half-lives
by several orders of magnitud. For example, for E$_{0}$=1.0 MeV, we have 
obtained within the  ATDHFB scheme 
t$_{SF}$= 9.504 $\times$ 10$^{35}$, 4.171 $\times$ 10$^{26}$ and 2.417 $\times$ 10$^{19}$ s
for $\eta$=1.00, 1.05 and 1.10, respectively. The corresponding GCM values turn out to be 
8.999 $\times$ 10$^{27}$, 8.440 $\times$ 10$^{20}$ and 1.831 $\times$ 10$^{15}$ s.

The spontaneous fission half-lives $t_{SF}$, predicted within the GCM and ATDHFB schemes, 
for the isotopes $^{232-280}$Pu  are 
 depicted, as functions of the neutron number, in Fig. \ref{tsf-eta-D1M}. 
 Results have been obtained with the
normal ($\eta$=1.00) and modified ($\eta$=1.05 and 1.10)
Gogny-D1M EDFs. Calculations have been carried out  with 
 $E_{0}$=0.5 [panel (a)], 1.0 [panel (b)], 1.5 [panel (c)] and 
 2.0 MeV [panel (d)], respectively. The experimental $t_{SF}$ values 
 \cite{Refs-barriers-other-nuclei-3-tsf}
 for  $^{236-244}$Pu
 are included in the plot. In addition,  $\alpha$-decay half-lives 
 \cite{TDong2005}
 are plotted
 with  short dashed lines. The results shown in Fig. \ref{tsf-eta-D1M} clearly 
 demonstrate, regardless of the ATDHFB and/or GCM scheme used, the strong impact of pairing
  correlations on the predicted t$_{SF}$ values. Note that, for example, our theoretical
  values for $^{236-244}$Pu agree reasonably well with the experimental ones. It is quite 
  satisfying to see that, in spite of the large variability in the 
  predicted t$_{SF}$ values, the main findings  previously summarized in Fig. \ref{tsf-D1M} 
still hold. On the one hand, these results and the ones discussed in Sec. \ref{FB-systematcis}
corroborate the predictive power of the Gogny-D1M EDF when used to describe fission along isotopic 
chains \cite{Rayner-Robledo-fission-U}. On the other hand, they also point to the use of 
experimental fission data to fine tune the pairing strengths in those EDFs commonly 
employed in fission calculations.

\section{Conclusions}
\label{conclusions}

In this paper we have considered, for the first time, the 
systematic 
microscopic description of fission
along the Plutonium  chain, including very neutron-rich isotopes, based on the 
Gogny-D1M EDF. In addition, we have further validated  
the use of the parametrization 
D1M to describe the fission properties of heavy nuclear systems 
through the computation of the spontaneous fission half-lives in $^{242-262}$Fm and their 
comparison with the available experimental data.

 We have resorted to 
the methodology already employed in Ref.\cite{Rayner-Robledo-fission-U} to determine
the fission paths 
(i.e., the 1F and 2F HFB solutions)
in $^{232-280}$Pu and $^{242-262}$Fm. In particular, we have considered 
constraints on the proton $\hat{Z}$ and neutron $\hat{N}$ numbers as well as 
on the (axially symmetric) quadrupole $\hat{Q}_{20}$, octupole $\hat{Q}_{30}$, necking 
$\hat{Q}_{Neck}(z_{0},C_{0})$
 and 
$\hat{Q}_{10}$ operators. Zero point rotational and vibrational quantum corrections have always been added to 
the corresponding 1F and 2F HFB configurations in an approximate projection-after-variation
(PAV)
framework \cite{rs}.

The spontaneous fission half-lives t$_{SF}$
for the considered nuclei,
have been computed within the standard WKB
approximation combined with microscopically determined state-of-the-art input resulting 
from the Gogny-D1M HFB calculations. The uncertainties 
 arising from such an input have been critically addressed. We have paid especial 
attention to the impact of pairing correlations on the spontaneous fission half-lives in 
$^{232-282}$Pu. Similar to the 
results obtained for the nuclei $^{232-282}$U 
\cite{Rayner-Robledo-fission-U}, we have found 
that changes of 5 $\%$ and 10 $\%$ in the pairing strengths of the original Gogny-D1M
EDF already lead to differences of several orders of magnitud in the theoretical t$_{SF}$ values. 
Let us stress that HFB calculations, based on the D1S and D1N Gogny-EDFs,  have also
been performed for the isotopes $^{232-282}$Pu. They reveal similar trends 
and variability
as the ones discussed 
in this study. From these results we conclude 
that, regarless of the particular parametrization of the Gogny-EDF employed, pairing 
correlations have a strong impact on  the absolute values of  fission observables
in Uranium, Plutonium and other heavy nuclei. This is further
corroborated from the recent results obtained with the 
Barcelona-Catania-Paris-Madrid (BCPM) EDF
\cite{Robledo-Giulliani}. Therefore, our calculations 
point to the use of fission data to fine tune the pairing 
strengths of those EDFs commonly employed in microscopic 
nuclear structure studies.

Nevertheless, in spite of the large variability observed 
in the results, a clear pattern emerges as a function of the mass number
in Uranium and Plutonium nuclei: the t$_{SF}$ values remain relatively constant 
up to N=166-168 and from there on they increase almost linearly up to 
a maximum  at the magic neutron number N=184. For increasing 
neutron number fission becomes faster than $\alpha$-decay.  For Plutonium isotopes, our 
calculations predict the crossing point to be around two mass units later (i.e., N $\approx$ 160)
than for Uranium isotopes. In addition, the 1F curves obtained for
 several Plutonium isotopes 
reveal that the shell effects responsible for the appearence of 
second fission isomers in this region of the nuclear 
chart are systematically present 
in ours and other mean-field calculations \cite{Mcdonell-2,Robledo-Giulliani,Rayner-Robledo-fission-U}. A detailed
investigation of those shells effects as well as the relation between 
second isomeric and dimolecular states in the framework of our calculations
is in progress and will
be reported elsewhere. 

In our calculations the masses and charges of the fission fragments are 
determined from 2F solutions obtained applying the Ritz variational 
principle to the HFB energy. Such an approximation, overestimates the role of the 
proton Z=50 and neutron N=82 magic numbers to determine the properties of the 
fission fragments. In particular, we have found a systematic overestimation 
of the heavy fragment's mass resulting from fissioning both Uranium and 
Plutonium nuclei. This indicates the need of a more sophisticated 
approximation than ours able to account for 
prescission quantum shell effects
as well as the broad mass distribution of the fission fragments 
\cite{Goutte-dynamical-distribution,Goriely-dist}.

Though in this and in our previous study
\cite{Rayner-Robledo-fission-U}
we
have shown that the Gogny-D1M HFB framework does provide a reasonable 
starting point, it is also clear that some of its defficiencies are 
deeply rooted in the description of fission resorting to minimal
energy criteria. Here, alternatives approaches
based on a minimal action, instead of a minimal energy, path 
\cite{deltaS-Doba}
deserve further consideration. Having in mind the strong impact
of pairing correlations on the predicted spontaneous fission half-lives
such theories should incorporate, in addition to the multipole moments, the 
minimization of the action Eq.(\ref{Action}) with respect to pairing 
fluctuations arising from the broken  U(1) number symmetry
in the intrinsic HFB states used to label the different fission 
configurations. Work along these lines is in progress and will 
be reported in a shortcoming publication.

\begin{acknowledgments}
Work supported in part by MICINN grants Nos. FPA2012-34694, FIS2012-34479 and by the 
Consolider-Ingenio 2010 program MULTIDARK CSD2009-00064. One of us (R.R), would 
like to thank the warm 
hospitality received at the Department of Physics, Kuwait University, during the 
first stages of this work.

\end{acknowledgments}

\end{document}